\documentclass[preprint,12pt,authoryear]{elsarticle}

\usepackage{amssymb}
\usepackage{amsmath}
\usepackage{graphicx}
\usepackage{color,soul}
\usepackage{hyperref}
\usepackage{gensymb}
\usepackage{relsize}

\usepackage{natbib}
\journal{Icarus}

\def\Rj{{\rm R_{jup}}}

\begin{document}

\begin{frontmatter}


\title{The effect of Jupiter oscillations on Juno gravity measurements}

\author[uniroma]{Daniele Durante\corref{corauthor}}
\ead{daniele.durante@uniroma1.it}
\cortext[corauthor]{Corresponding author}

\author[oca]{Tristan Guillot}
\ead{tristan.guillot@oca.eu}

\author[uniroma]{Luciano Iess}
\ead{luciano.iess@uniroma1.it}

\address[uniroma]{Sapienza University of Rome, Dipartimento di Ingegneria Meccanica ed Aerospaziale, Via Eudossiana 18, 00184 Rome, Italy}
\address[oca]{Universit\'e de Nice-Sophia Antipolis, Observatoire de la
  C\^ote d'Azur, CNRS UMR 7293, 06304 Nice Cedex 4, France}


\begin{abstract}
Seismology represents a unique method to probe the interiors of giant planets. Recently, Saturn's f-modes have been indirectly observed in its rings, and there is strong evidence for the detection of Jupiter global modes by means of ground-based, spatially-resolved, velocimetry measurements. We propose to exploit Juno's extremely accurate radio science data by looking at the gravity perturbations that Jupiter's acoustic modes would produce. 
We evaluate the perturbation to Jupiter's gravitational field using the oscillation spectrum of a polytrope with index 1 and the corresponding radial eigenfunctions. We show that Juno will be most sensitive to the fundamental mode ($n=0$), unless its amplitude is smaller than 0.5 cm/s, i.e. 100 times weaker than the $n \sim\ 4 - 11$ modes detected by spatially-resolved velocimetry. The oscillations yield contributions to Juno's measured gravitational coefficients similar to or larger than those expected from shallow zonal winds (extending to depths less than 300 km). In the case of a strong f-mode (radial velocity $\sim$ 30 cm/s), these contributions would become of the same order as those expected from deep zonal winds (extending to 3000 km), especially on the low degree zonal harmonics, therefore requiring a new approach to the analysis of Juno data. 
\end{abstract}


\begin{keyword}
Geophysics \sep Jupiter interior \sep Orbit determination
\end{keyword}

\end{frontmatter}


\section{Introduction}
\label{sec:1}

Arriving at Jupiter on 4 July 2016, NASA's Juno mission will complete 37 orbits around the planet, revealing details of its interior structure and composition \citep{Bolton2010}. A radio science experiment will measure Jupiter's gravity field with extremely high accuracy, thereby providing constraints on the planet's interior density profile and differential rotation \citep{Hubbard1999, Guillot2005, Kaspi2010}. Juno's orbit is polar and highly eccentric (e=0.95), with a period of about 14 days. The perijove altitude is about 4000 km above the reference 1 bar level (which corresponds to 71492 km at the equator). The pericenter latitude precesses from 5\degree N to 35\degree N over the nominal one-year and a half mission. The longitude of the node is controlled by means of orbital maneuvers. During pericenter passes Juno is tracked from ground at Ka band (32-34 GHz) to obtain the spacecraft range rate to accuracies of a few micron/s over time scales of 1000 s \citep[see][]{Asmar2005}. The low pericenter altitude and the very accurate radio system provide excellent sensitivities to the gravity field of the planet \citep{Finocchiaro2010, Finocchiaro2013}.

Another method to probe the interior of the planet is through the determination of Jupiter's acoustic normal modes. While these oscillations are certainly a potential source of information on the radial density profile, they may also complicate the interpretation of Juno's gravity data. Jupiter's normal modes displace large masses that may perturb the spacecraft motion to levels that can be measured by Juno's extremely accurate Doppler system.

Theoretical studies of Jupiter's seismology were first performed in the mid 1970s \citep{Vorontsov1976}. They were subsequently revised and extended to include the effect of Jupiter's rotation and flattening \citep{Vorontsov1981, Mosser1990}, atmosphere \citep{Mosser1995}, and to infer consequences for Jupiter's interior models \citep{Gudkova1999, Jackiewicz2012}.

Over the past decades, several attempts to observe Jovian global modes have been carried out with different methods: thermal infrared photometry \citep{Deming1989}, optical resonance spectrophotometry \citep{Schmider1991}, using a Fourier transform spectrometer \citep{Mosser1993}, and finally with SYMPA (Seismographic Imaging Interferometer for Monitoring of Planetary Atmospheres), a Fourier tachometer whose principle is based on the spectro-imaging of the full planetary disk \citep{Schmider2007}. Analyzing data acquired with the latter instrument, \cite{Gaulme2011} detected an excess power between 800 and 2100 $\mu$Hz and a secondary excess power between 2400 and 3400 $\mu$Hz, as well as a characteristic splitting of the peaks of 155.3 $\pm$ 2.2 $\mu$Hz, all of these compatible with frequencies of acoustic oscillations predicted by interior models of Jupiter.

In parallel, \cite{Hedman2013} confirmed that waves observed by the Cassini spacecraft in Saturn's rings cannot be caused by satellites and therefore must result from global oscillations in Saturn, as had been proposed by \cite{Marley1993}. In this case, the very accurate determination of mode frequencies allows to directly probe the interior structure. The modes observed are f-modes, i.e., they correspond to the fundamental ($n=0$) mode of acoustic oscillations. \cite{Fuller2014} proposed that the observed frequency splitting results from a mixing mechanism between f-modes and g-modes (i.e., non-acoustic waves for which the restoring force is gravity), which requires the existence of a deep thick stable stratified region within the interior of Saturn.

It is thus clear that both Jupiter and Saturn oscillate (to some degree) and that a large fraction of the mass of these planets is involved in these oscillations. We aim to determine whether these oscillations have consequences for Juno, both for our ability to properly estimate the gravitational moments and then to provide constraints on the characteristics of those oscillations. 

The article is organized as follows: In section \ref{sec:2}, we describe the simplified acoustic oscillation spectrum of Jupiter adopted as a basis for our calculations.  In section \ref{sec:3}, we explain how we model gravitational perturbations arising from the oscillations. The consequences for Juno's measurements are presented in section \ref{sec:4}. The last section summarizes our findings. 


\section{Jupiter's acoustic oscillation spectrum}
\label{sec:2}

As first demonstrated by \cite{Vorontsov1976}, Jupiter's atmosphere reflects acoustic waves if their frequency is smaller than about 3 mHz. These waves may propagate into the planetary interior, down to the core for some of them. At some well-defined frequencies, these waves resonate constructively and can reach non-negligible amplitudes. These frequencies depend on the structure of the planet. As long as a precise determination of the wave frequencies is not required, we estimate them by using simple models. The amplitudes of the waves depend on several unknown factors: the structure of the planet but also excitation and damping mechanisms. At this stage it is hard to identify a clear excitation mechanism. In terms of damping in the absence of other mechanisms yet to be determined, radiative damping in the atmosphere will yield a slow decay of waves with a timescale of about 300 years at 1 mHz to 10 days at 3 mHz (see \citealp{Mosser1995} and \citealp{Gaulme2015}).  

\begin{figure}[t]
\centering
\includegraphics[height=2.5in]{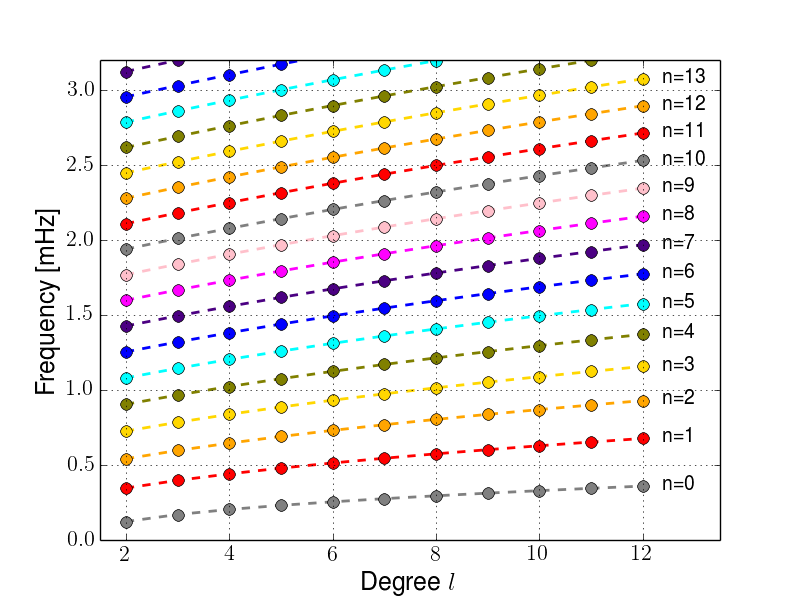}
\caption{Frequencies for low degrees and low radial order acoustic modes.}
\label{fig:1}
\end{figure}

Following \cite{Hubbard1977} (see also \citealp{Guillot2005}), we choose to approximate Jupiter's interior structure by a simple polytropic model of polytropic index equals to 1. The simplicity of this model does not compromise the results and the conclusions of this paper. The spectrum of mode frequencies surely differs from those predicted by more sophisticated model, but for our purposes, slightly different frequency values do not change substantially the overall picture of the gravitational signature of Jupiter acoustic modes. The same is true if the radial eigenfunctions are slightly changed. Indeed, our goal is simply an assessment of the effect of normal modes on Juno's gravity measurements. As a reference, frequencies obtained with the polytrope for low-degree, low-radial-order modes differs by about 5\% with respect to those reported by detailed models \citep[e.g.,][]{Gudkova1999}. From a geophysical point of view this aspect is crucial, but not for the purpose of this paper. Eigenfrequencies and eigenfunctions can then be easily computed for the polytropic model. As we comment later in section \ref{sec:31}, gravity measurements does not allow us to consider modes of degree 0 and 1. The frequencies of the first few low degree, low-radial-order modes are reported in fig. \ref{fig:1}, starting from $l=2$.

The period of the modes plays a crucial role in Juno's measurements. Due to the high eccentricity of Juno's orbit, the measurement sensitivity to the harmonics of gravity field is higher near perijove, and decreases sharply as the spacecraft altitude increases (gravity measurements are carried out at about +/- 4 hours, corresponding to $\sim 5.8\,\Rj$, from closest approach). Fundamental modes have periods of about one hour, to a maximum value of 2 hours for the $l=2$ mode. Higher radial order modes have progressively shorter periods. The modes detected by SYMPA have periods between about 5 to 20 minutes and would correspond to $n \sim 4$ to $11$ modes for l=2 modes. Therefore the timescale of these phenomena falls within Juno's sensitivity window, set by the duration of a perijove pass (8 hours).

\begin{figure}[t]
\centering
\includegraphics[height=2.5in]{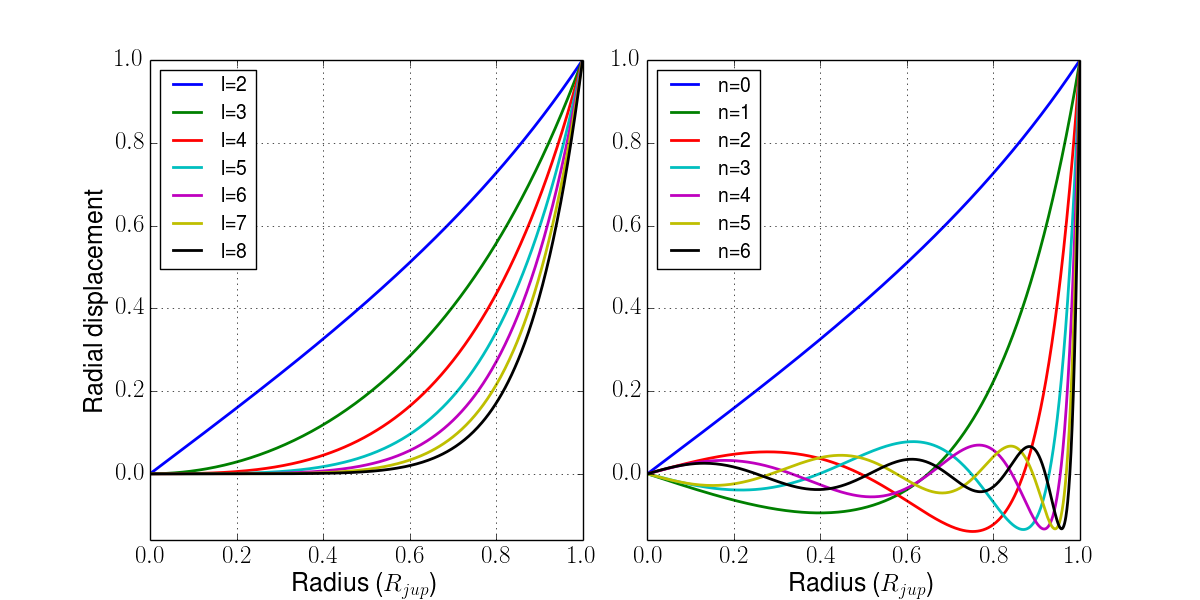}
\caption{Radial eigenfunctions versus distance (Jovian radii), normalized at the surface. On the left, fundamental modes with increasing degree; on the right, modes with $l=2$ and increasing radial order.}
\label{fig:2}
\end{figure}

Jupiter radial eigenfunctions are reported in fig. \ref{fig:2}. On the left, eigenfunctions for fundamental modes ($n=0$), affecting the whole planet, are reported for $m=0$ and increasing degree, whereas modes with degree $l=2$, $m=0$, and increasing radial order are shown on the right. Eigenfunctions are normalized to be unity at Jupiter reference radius.

We can notice on the left panel of fig. \ref{fig:2} that, as the degree of the spherical harmonic increases, the fundamental modes move towards the surface, and the perturbation influences a limited portion of the planet (the outer layers). Similarly (right panel on fig. \ref{fig:2}), when the radial order increases, the oscillation period decreases, and the corresponding eigenfunction reaches values close to unity only near the surface. High radial order oscillation modes influence the Jupiter upper layers, whereas low radial order modes penetrate deeper in the planet. It is a very well known aspect of acoustic modes: they probe different region of the interior of a planet (or a star) according to its degree and radial order.


\section{Gravity perturbation modeling}
\label{sec:3}

\subsection{Basics}
\label{sec:31}

In order to evaluate the perturbation to Jupiter gravitational field produced by acoustic modes, we start from the harmonic expansion of the gravitational potential:

\begin{equation}
\label{eq:1}
U(r,\theta,\varphi)=-\frac{GM}{r} \left\lbrace 1 + \sum_\mathsmaller{l \geq 2} \sum_\mathsmaller{\ -l \leq m \leq l}{\left[ \left(\frac{R}{r}\right)^l \ U_{l,m} \ Y_{l,m}(\theta,\varphi) \right] } \right\rbrace
\end{equation}

With the usual convection, $R$ is the reference Jupiter radius, $r$ is the radial coordinate, $\theta$ is the colatitude, and $\varphi$ is the longitude at which the potential is computed. The term $-\frac{GM}{r}$ represent the monopole term, $l=0$ in the harmonic expansion. In addition, by imposing terms with $l=1$ to zero, the center of mass is found at the origin of the reference frame. The normalized spherical harmonic of degree $l$ and order $m$ is denoted by $Y_{l,m} (\theta, \varphi)$.

The internal density distribution $\rho(r)$ determines the spherical harmonic coefficients $U_{l,m}$ through \citep{Bertotti2003}:

\begin{equation}
U_{l,m}=\frac{\int_{V}{r'^l\ Y_{l,m}(\theta',\varphi')\ \rho(P')} \, dV}{(2l+1)\ M\ R^l}
\end{equation}

The integral is over the volume of the reference sphere, with the volume element $dV=r'^2\ sin{\theta'}\ dr'\ d\theta'\ d\varphi'$. The density is computed at the internal point $P'$, normalized by the total mass of the planet, $M$.

The density profile of Jupiter is perturbed by acoustic oscillation, thus:

\begin{equation}
\rho(r',\theta',\varphi',t)=\overline{\rho}(r',\theta',\varphi')+\sum_\mathsmaller{l \geq 2} \sum_\mathsmaller{\ -l \leq m \leq l} \sum_\mathsmaller{\ n \geq 0}{\left[ \Delta\rho_{l,m,n}(r',\theta',\varphi',t) \right]}
\end{equation}

The internal unperturbed density profile $\overline{\rho}(r',\theta',\varphi')$ is disturbed by the superposition of perturbation characterized by different degree and orders, $\Delta\rho_{l,m,n} (r',\theta',\varphi',t)$. Each of these density perturbations must conserve the total amount of mass, and its temporal average over long time-scale must be zero. Thus, they must satisfy:

\begin{equation}
\label{eq:4}
\int_{V}{\Delta\rho_{l,m,n}(r',\theta',\varphi',t)} \, dV = 0
\end{equation}

\begin{equation}
\label{eq:5}
\overline{\Delta\rho_{l,m,n}}(r',\theta',\varphi',t) = 0
\end{equation}

The last two equations must be satisfied for any density perturbation associated to all allowable degree and order. We can write the density perturbation as:

\begin{equation}
\Delta\rho_{l,m,n}(r',\theta',\varphi',t) = \widetilde{\Delta\rho}_{l,m,n}(r',\theta',\varphi') \ \cos( \omega_{l,m,n} t + \phi_{l,m,n} )
\end{equation}

The term $\widetilde{\Delta\rho}_{l,m,n} (r',\theta',\varphi')$ indicates the maximum amplitude of the perturbation. In the cosine term, $\omega_{l,m,n}$ is the mode frequency and $\phi_{l,m,n}$ accounts for the oscillation phase. The first factor is:

\begin{equation}
\label{eq:7}
\widetilde{\Delta\rho}_{l,m,n}(r',\theta',\varphi') = \left . \left(\frac{\partial \rho}{\partial r'}\right) \right |_{r'} \left[ A_{l,m,n} \ f_{l,m,n}(r') \ Y_{l,m}(\theta',\varphi') \right] 
\end{equation}

In equation~\eqref{eq:7}, $\left . \left(\frac{\partial \rho}{\partial r'}\right) \right |_{r'}$ is the density gradient at a given internal radius $r’$, $f_{l,m,n} (r')$ is the radial eigenfunction associated to the acoustic mode of degree $l$, azimuthal order $m$, and radial order $n$, whereas $A_{l,m,n}$ is the displacement of the upper troposphere.

By construction, the proposed formulation automatically satisfies the relation given in eq.~\eqref{eq:5} for every admissible value of $l$, $m$, and $n$. On the other hand, the condition of mass conservation (eq.~\ref{eq:4}), is not met when $l=0$. This is because the proposed formulation does not account for the expansion and contraction of Jupiter' shape. In addition, a $l=0$ density perturbation (of arbitrary order $n$) does not affect external gravity, because the external gravitational potential is invariant to spherically symmetric variations of the internal density distribution. Furthermore, the relation given in eq.~\eqref{eq:7} should not be applied to the dipole $l=1$ terms. Indeed, a perturbation associated to a dipole potential would produce a displacement of the center of mass of the planet, which would violate conservation of momentum. Thus, eq.~\eqref{eq:7} only applies to modes with $l\geq2$ and arbitrary order $m$ and $n$.

The above formulation allows the computation of the dynamic contribution to Jupiter gravitational potential. The spherical harmonic coefficients of the gravity field expansion given in eq.~\eqref{eq:1} include now the dynamic part due to the acoustic modes:

\begin{equation}
\label{eq:8}
U_{l,m} = U_{l,m}^\mathsmaller{STATIC} + \sum_\mathsmaller{n \geq 0} \widetilde{U}_{l,m,n} \ \cos( \omega_{l,m,n} t + \phi_{l,m,n} )
\end{equation}

The dynamic spherical harmonic coefficients (identified by an upper tilde) are:

\begin{equation}
\label{eq:9}
\widetilde{U}_{l,m,n}=\frac{\int_{V}{r'^l\ Y_{l,m}(\theta',\varphi')\ \widetilde{\Delta\rho}_{l,m,n}(P')} \, dV}{(2l+1)\ M\ R^l}
\end{equation}

The last formula allows the computation of acoustic perturbations to the gravity field coefficients. The integral can be computed numerically in spherical coordinates. Spherical harmonic functions are well defined in the literature, and the density perturbation at a given point P' is given by eq.~\eqref{eq:7} (Jupiter mass and reference radius are well known). The amplitudes of the coefficients depend only by the free parameters $A_{l,m,n}$  , i.e., the surface displacement for the different modes.
The gravitational potential, including the perturbation from Jupiter modes, writes:

\begin{equation}
\begin{split}
U & (r,\theta,\varphi) = -\frac{GM}{r} \left\lbrace 1 + \sum_\mathsmaller{l \geq 2} \sum_\mathsmaller{-l \leq m \leq l}{\left[ \left(\frac{R}{r}\right)^l \ U_{l,m} \ Y_{l,m}(\theta,\varphi) \right]} + \right . \\
& \left . + \sum_\mathsmaller{l \geq 2} \sum_\mathsmaller{-l \leq m \leq l} \sum_\mathsmaller{n \geq 0}{\left[ \left(\frac{R}{r}\right)^l \  \widetilde{U}_{l,m,n} \ Y_{l,m}(\theta,\varphi) \ \cos( \omega_{l,m,n} t + \phi_{l,m,n} ) \right] } \right\rbrace
\end{split}
\end{equation}

The gravitational acceleration equals the gradient of this gravitational potential:

\begin{equation}
\label{eq:11}
a^{BF} = - \nabla U
\end{equation}

The superscript BF indicates that the acceleration is computed in the Jupiter body fixed reference frame. The components of the acceleration (eq.~\ref{eq:11}) in the inertial reference frame (required for the integration of the trajectory) are evaluated using the IAU rotation model for Jupiter \citep{Archinal2010}.

\subsection{Modes amplitudes and perturbed harmonic coefficients}
\label{sec:3.2}

At present, the only evidence of large acoustic oscillations on Jupiter comes from the analysis of radial velocity maps of Jupiter’s upper troposphere. The results provided by \cite{Gaulme2011} suggest a mode peak amplitude, in term of radial velocity, of about:

\begin{equation}
v \simeq 49 \ \rm cm/s
\end{equation}

The frequency of maximum amplitude in the first window (800-2100 $\mu$Hz) has been found to be:

\begin{equation}
f \simeq 1213 \pm 50 \ \rm \mu Hz
\end{equation}

These results are in broad agreement with the theoretical values \citep{Vorontsov1976, Bercovici1987, Mosser1995} and previous observations \citep{Mosser2000}. However the mechanism responsible for exciting these waves at these high amplitudes remains unknown (e.g., \citealp{Gaulme2015}).

\cite{Fuller2014} estimates that the mode amplitudes detected in Saturn by \cite{Hedman2013} are about 1000 times smaller than those found by \cite{Gaulme2011}. This difference is in line with what has been observed on the Sun, where the amplitudes of low frequency modes are generally much smaller than the peak amplitude at higher frequency modes \citep{Goldreich1994}.

We would like to stress that Jupiter's detected modes are essentially p-modes (with radial order $n \sim 4-11$ and a low degree $l\sim1$), whereas for Saturn only f-modes ($n=0$) have been observed. The main difference between the two is essentially the associated frequency: f-modes have lower frequency than p-modes. For the Sun, the ratio between the observed mean velocity associated to p-modes and the observed mean velocity associated to f-modes (equal degree) is as order of magnitude $\sim 100$ [J.~Jackiewicz, personal communication]. Thus, we may expect p-mode velocities to be larger than f-modes also for Jupiter and Saturn.

Unfortunately, \cite{Gaulme2011} were not able to identify the degree and order of the spherical harmonics expansion of Jupiter's modes due to the high correlation between the projection of spherical harmonics into 2-dimensions, mainly produced by the resize of the sensitivity area to 75\% of the Jovian diameter. For this reason, simplifying assumptions have to be made about the mode amplitudes:

\begin{enumerate}
\item We suppose that the radial velocity associated to each mode depends only by its oscillation frequency, and not directly by its degree or radial order. This is in agreement with the observation of solar p-modes [J.~Jackiewicz, personal communication];
\item Although the excitation mechanisms certainly do not privilege zonal modes, Juno is mostly sensitive to a zonal field. We therefore focus on the effects of those components of acoustic oscillations;
\item We assume a Gaussian profile for the radial velocity to frequency mapping function, plus a constant radial velocity, independent of modes characteristics.
\end{enumerate}

With these assumptions, each mode's amplitude depends only by its frequency, and the mapping function is:

\begin{equation}
v(f) = v_{bias} + (v_{max}-v_{bias}) \exp{\left[ -\frac{1}{2} {\left( \frac{f-f_{max}}{\sigma_{f}} \right)}^2 \right]}
\end{equation}

The velocity-to-frequency profile is constructed to match the observed peak frequency (1210 $\mu$Hz), and the maximum observed radial velocity (50 cm/s). The parameter $\sigma_f$ is the standard deviation in frequency, which is related to the Gaussian distribution assumed for the radial velocity. The quantity $v_{bias}$ is the additive constant value, independent of the mode's frequency. To compute the surface displacement, we divide each mode’s mean radial velocity at the surface by its frequency:

\begin{equation}
\left . A_{l,m,n} \right|_{r=R} = \frac{v(f_{l,m,n})}{2 \pi f_{l,m,n}}
\end{equation}

The surface displacement allows the computation of the perturbation on harmonic coefficients. However, because we have information only within the observed frequency range, the quantities $v_{bias}$ and $\sigma_f$ are unknown, with no direct indication of their putative value. We explore two different scenarios to account for the uncertainty in the velocity-to-frequency profile, identified as case \textit{A} and case \textit{B}.

Case \textit{A} is a nominal case. The value for $v_{bias}$ is selected to match (in order of magnitude) the observed ratio between the amplitudes of p-modes and the amplitude of the f-modes in the Sun. This ratio represents the relative contribution to p-modes with respect to f-modes. For the Sun, the maximum ratio is about 100 (it changes with the degree considered). For Jupiter, we can achieve this by imposing $v_{bias}=0.3$ cm/s, and selecting the other free parameter to match the trend in the observed values of the mean radial velocity around the maximum amplitude frequency, thus setting $\sigma_f=300$~$\mu$Hz. The fact that our function is a relatively poor approximation of the spectral amplitudes at high frequencies is inconsequential: as we will see afterwards, high order modes have much smaller effects on the global gravitational acceleration.

\begin{figure}[h]
\centering
\includegraphics[height=2.5in]{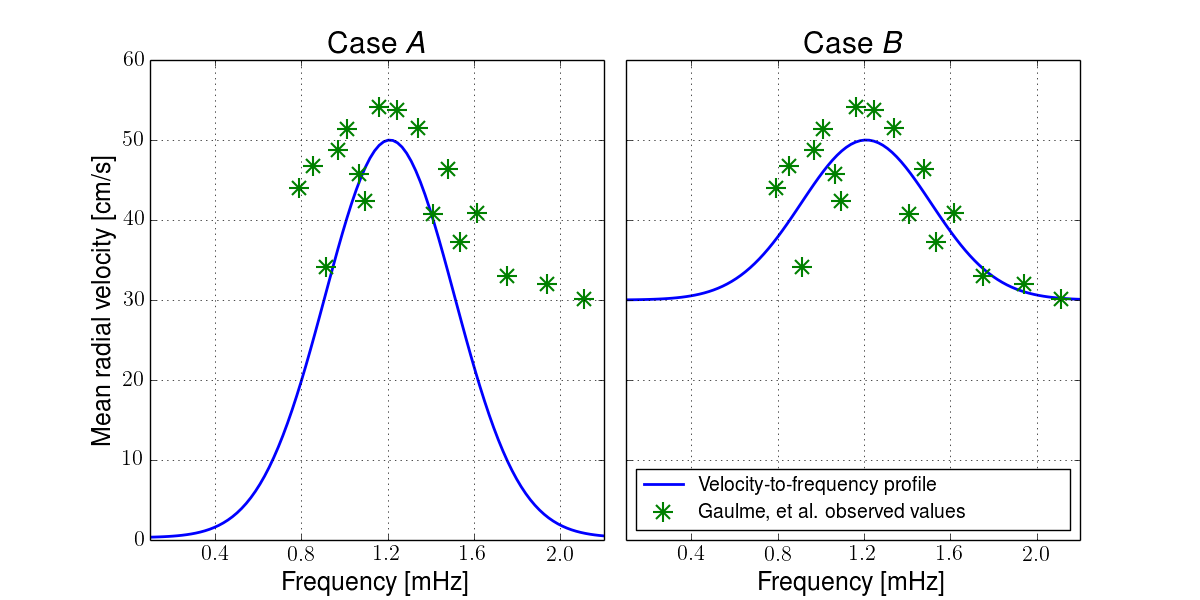}
\caption{Mean radial velocity versus frequency for two selected cases, compared to the observed value reported by \cite{Gaulme2011}. Note that the observations do not offer indications on the mode amplitude at low frequency. The two Gaussian profiles are attempts to extrapolate the mode amplitude outside the observed frequency range. }
\label{fig:3}
\end{figure}

The second scenario, case \textit{B}, depicts a very energetic Jupiter, with large f-mode amplitudes. The bias velocity is selected as to match SYMPA's observed values for radial velocity at high frequencies (larger than the peak amplitude frequency), neglecting the contribution from the background noise. Thus, we set $v_{bias}=30$ cm/s, and $\sigma_f=300$ $\mu$Hz. These values correspond to a profile that fits the observed values reported in \cite{Gaulme2011}. Fig. \ref{fig:3} shows the velocity-to-frequency profiles for the two cases considered, with observed values reported for comparison. This unphysical case represents an upper limit set by current observation.

\subsection{Consequences for Jupiter's static gravity field}

In order to evaluate the contribution to Jupiter gravity field coming from acoustic modes in the two cases, we compare the acoustic gravity harmonics with those expected from static gravity. Jupiter gravity field is supposed to be the superposition of two main contributions: Jupiter solid body rotation, described by \cite{Hubbard2012}, and wind dynamics, which contribute (according to their penetration depth) to perturb the gravity field as described by \cite{Kaspi2010}. 

\begin{figure}[h]
\centering
\includegraphics[height=2.5in]{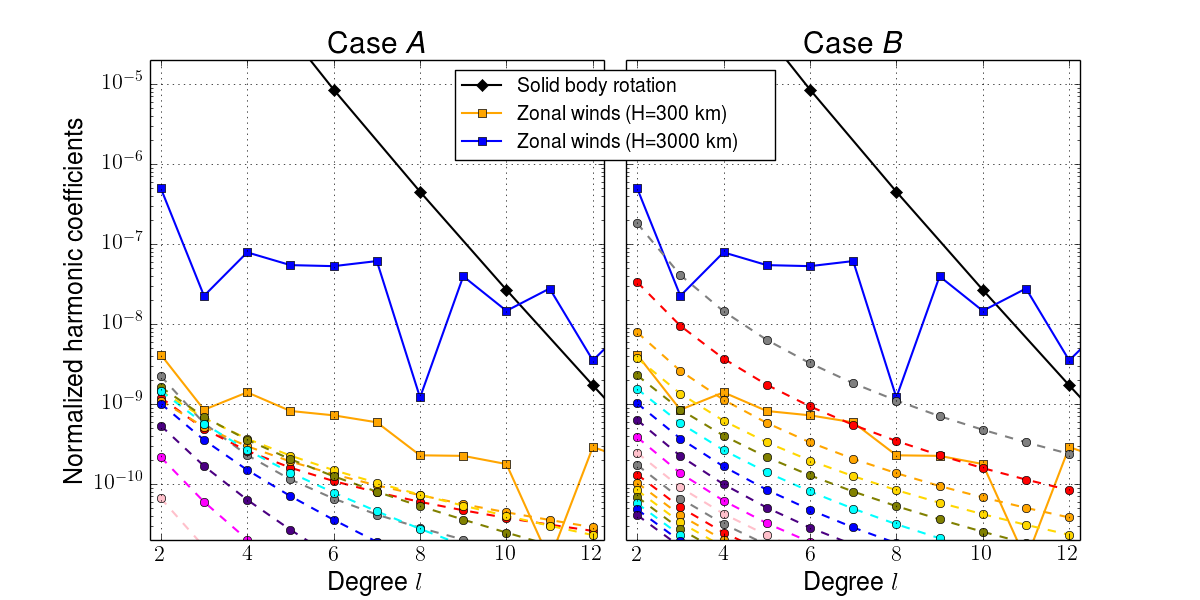}
\caption{Perturbation to normalized harmonic coefficients of Jupiter's gravity field due to acoustic modes (colors), compared to the solid body rotation (black) and possible zonal wind contributions (blue and yellow, as labeled). The colors for the acoustic modes correspond to different radial orders, as in fig. 1 (e.g., grey for $n=0$, red for $n=1$, orange for $n=2$, yellow for $n=3$, etc.).}
\label{fig:4}
\end{figure}

A summary plot of the harmonic coefficients for the selected cases is reported in fig. \ref{fig:4}, including modes with different degree and radial order. For comparison, we report also the static contribution due to zonal winds for two different penetration depths: $H=300$ km (shallow winds), and $H=3000$ km (deep winds).

Figure \ref{fig:4} shows that the expected dynamical contribution to gravity field harmonics tends to decrease for increasing radial order (for $n>8$) and increasing harmonic degree. Thus, low degree, low radial-order modes give the largest contribution to the gravity signal. Note that high order mode amplitudes ($n\simeq5-7$) are essentially fixed by the maximum velocity given by the observations, whereas for the $n=0$ f-modes (and generally for low $n$ modes) we do not have direct indications, and the amplitude depends substantially on the parameter $v_{bias}$. Indeed, in fig. \ref{fig:3} we notice that in case \textit{B} a large bias exists for all modes, while in case \textit{A} the low frequency modes have a much smaller amplitude.

We also note that in case \textit{A} the perturbing coefficients have about the same order of magnitude of shallow winds coefficients. In case \textit{B}, gravitational perturbations due to acoustic modes are very significant, with magnitudes comparable to the contribution coming from density perturbations due to wind circulation in a deep zonal winds scenario. In the next section, we estimate the consequences of oscillations for Juno's dynamical measurements of Jupiter's gravity field.


\section{Juno sensitivity to dynamical gravitational field}
\label{sec:4}

\subsection{Juno's gravity experiment}
\label{sec:4.1}

Juno's gravity experiment aims to estimate Jupiter's gravity field through the analysis of radio tracking data. The crucial observable quantity is the Doppler shift of a two-way, coherent, radio link established in Ka-band (32-34 GHz) between a ground station and the spacecraft. In that configuration, the ground station transmits a reference radio signal to the spacecraft, which retransmit the incoming signal coherently in phase to the same ground station on Earth. Doppler data provide the spacecraft range rate (relative velocity projected on the line-of-sight), enabling the reconstruction of spacecraft trajectory as well as the estimation of the gravity moments. The onboard Ka-band translator system (KaTS) is the key element of the radio system, enabling range rate measurements accurate to 20 $\mu$m/s at 60 seconds of integration time \cite[see][for a complete discussion]{Finocchiaro2010}. Such an accuracy allows in principle to sample accelerations as low as $3 \times 10^{-10}$ km/s\textsuperscript{2}. However, Juno's real performance may be accurately evaluated just by means of numerical simulations that include the processing of synthetic Doppler data in a precise Orbit Determination (OD) software.

Previous studies (see \citealp{Finocchiaro2010}, \citealp{Finocchiaro2013}, and \citealp{Tommei2015}) assessed the main acceleration terms contributing to spacecraft motion near perijove (see table \ref{tab:1}). The main term is due to Jupiter's monopole ($2.3 \times 10^{-2}$ km/s\textsuperscript{2}), whereas the planet oblateness produces an effect one order of magnitude smaller. Higher order harmonics produce a smaller, but still well detectable, acceleration on the Juno spacecraft. The tide induced on Jupiter by its moon Io produces an acceleration on the spacecraft of about $10^{-8}$ km/s\textsuperscript{2}, again large enough to be measured by Juno's accurate Doppler system.

\begin{table}
\centering
\begin{tabular}{lc}
\hline
Cause & Value(km/s\textsuperscript{2}) \\
\hline
Jupiter monopole & $2.3 \times 10^{-2}$ \\
Jupiter oblateness ($J_2$) & $1.2 \times 10^{-3}$ \\
Jupiter $J_4$ & $5.0 \times 10^{-5}$ \\
Jupiter $J_6$ & $3.5 \times 10^{-6}$ \\
Solid tide (Io) & $8.9 \times 10^{-9}$ \\
\hline
\end{tabular}
\caption{Acceleration terms acting on Juno at pericenter, from \cite{Tommei2015}.}
\label{tab:1}
\end{table}

\begin{figure}[h]
\centering
\includegraphics[height=2.5in]{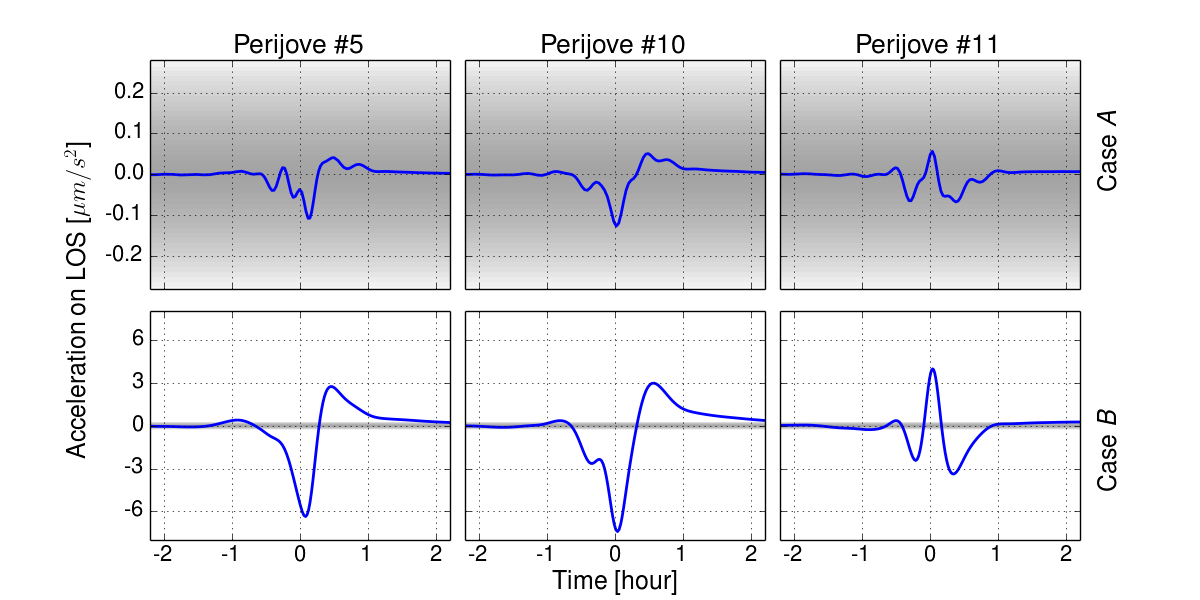}
\caption{Acceleration term caused on Juno by the acoustic modes, for the first three gravity orbits, in the selected cases. The time axis is centered about the perijove epoch.}
\label{fig:5}
\end{figure}

Fig. \ref{fig:5} shows the perturbing acceleration acting on Juno induced by acoustic oscillations, in the first three gravity orbits (\#5, \#10, and \#11), for case \textit{A} and \textit{B}. The peak acceleration occurs near perijove, for a total duration of about two hours. The gravitational signal changes with each orbit because it depends on how the different modes overlap, leading to an unpredictable initial phase of the modes at the beginning of each tracking arc. In our calculations, the phases of the modes are the same in the two cases but their amplitudes are very different, yielding different perturbing accelerations. We seek to evaluate when the signal becomes large enough to be seen in Juno data.

Juno's 20 $\mu$m/s Doppler precision at 60 s is close to the limit of current microwave systems. Case \textit{A} shows signals which are slightly smaller than this value, whereas in case \textit{B} they are more than an order of magnitude above it. However, the expected performance of the experiment depends on the number of solve-for parameters and the information matrix of the problem. It turns out that the static gravity field coefficients show significant mutual correlation \citep{Finocchiaro2010, Finocchiaro2013}. For this reason, we assessed Juno's capability by means of a full set of simulation with an orbit determination (OD) code.

\subsection{Orbit determination code and dynamical models}
\label{sec:4.2}

The setup used to investigate the performance of Juno radio science experiment includes two successive steps: data simulation and data estimation. Both of these steps have been carried out JPL's latest OD code, MONTE (Mission Analysis, Operations, and Navigation Toolkit Environment, see \citealp{Evans2016}), and a separate, multi-arc estimation filter for the estimation process \citep{Milani2009}.

Juno's synthetic Doppler data (two way, Ka-band radio observables) are simulated for all perijove arcs devoted to radio science passes. The initial conditions for the spacecraft (position and velocity) are extracted from the most up-to-date reference trajectory delivered by the Juno navigation team (spk\_ref\_150326\_180221\_150326.bsp, available at NASA's NAIF website). The dynamical model for the spacecraft, discussed in detail by \cite{Finocchiaro2010}, \cite{Finocchiaro2013}, and \cite{Tommei2015}, accounts for gravitational accelerations from all the solar system's bodies and from the Galilean satellites. The Doppler observable is computed in a relativistic context to the post-Newtonian first order correction. The solar pressure on Juno's large solar panels is included, as well as the acceleration due to Jupiter albedo and thermal emission. The tides raised on Jupiter by the Galilean satellites (the main contribution coming from Io) are included by means of Love numbers \cite[see][]{Murray1999}. Jupiter's gravity field has been simulated in the shallow winds case (penetration depth of 300 km) for the zonal winds \citep{Kaspi2010}. A full set of gravity field coefficients up to degree and order 30 has been included and Jupiter’s rotation model agrees with IAU convention \citep{Archinal2010}. To simplify the setup, the relativistic Lense-Thirring precession caused by the rapid rotation of Jupiter is not accounted for (the resulting acceleration on Juno would be of the order of $10^{-10}$ km/s\textsuperscript{2} at perijove). 

With such a scenario, Juno trajectories are integrated for a 24-hour arc centered at perijove for all 26 orbits devoted to gravity science. For each arc, synthetic two-way Ka-Band Doppler data, sampled at 60 sec of integration time, between Juno and NASA Deep Space Station 25 in Goldstone (DSS 25) are produced when the elevation angle is higher than 15 degree. Gaussian white noise is then added to the observables to simulate a realistic dataset.

The estimation process starts from the synthetic data set. The Juno state vectors at the beginning of each arc are estimated as local parameters. The global parameters (i.e. parameters that are common to all arcs) to be estimated are Jupiter gravitational parameter (GM), gravity field coefficients ($J_{2}$, $C_{21}$, $S_{21}$, $C_{22}$, $S_{22}$, and higher degree zonal coefficients up to degree 15), and degree 2 and 3 Love numbers. The estimation of this set of parameters allows us to fit the data to noise level when acoustic oscillations are not included \citep{Finocchiaro2010, Finocchiaro2013} and the winds are shallow. (In the deep wind case, a full 20x20 tesseral field is needed.) To obtain the estimate of global and local parameters, partial derivatives of the Doppler observables with respect to each parameter are computed on the reference trajectory, and Doppler residuals (observed minus computed) are evaluated. A multi-arc least square information filter gives the differential correction on estimated parameters and the covariance matrix of the problem. The process is iterated up to convergence.
In order to assess the influence of Jupiter's acoustic modes on the gravity field estimation, we have generated synthetic Doppler data that include the acoustic mode perturbations according to the model described in sect. \ref{sec:3.2}. We have then fitted the Doppler data using the standard dynamical model, where acoustic oscillations are not adjusted, nor included. Three cases shall be considered: 1) The modes do not produce perturbations large enough to affect neither the residuals nor the estimate; 2) The modes do not produce signatures on Doppler, but the estimated values of the static gravity field and Love numbers is biased; 3) The modes are so large to produce signatures in the Doppler residuals, which are not compatible with the expected noise level.

\subsection{Doppler residual analysis}

\begin{figure}[h]
\centering
\includegraphics[width=\hsize]{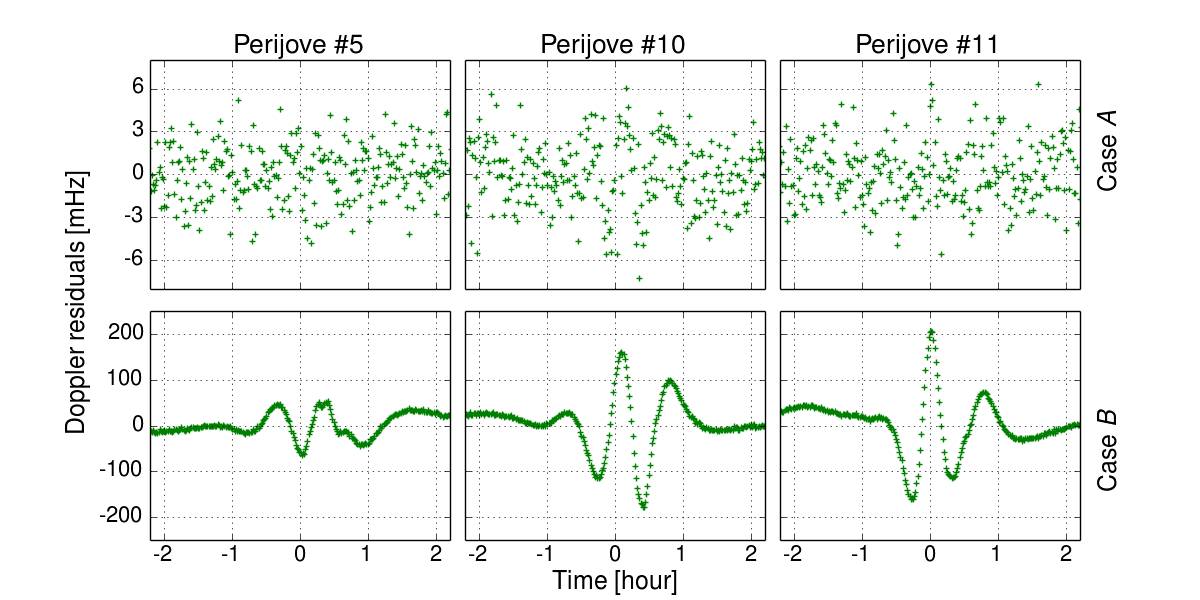}
\caption{Doppler residuals after a filtering attempt, for selected orbits and different cases. In case \textit{A} the filter produced zero-mean, nearly white noise residuals, whereas in case \textit{B} it does not.}
\label{fig:6}
\end{figure}

Inspection of Doppler residuals provides a first indication of the presence (or absence) of acoustic modes. Fig. \ref{fig:6} shows the post-fit Doppler residuals obtained in case \textit{A} and \textit{B}, for the first three gravity orbits (perijoves \#5, \#10, and \#11). The data are fitted to the noise level if the statistics of the residuals is compatible with the noise added in the simulation of the observables. In this paper we have considered only white frequency noise. Although in the real case other noise sources may complicate the error budget, the statistics of Ka band Doppler data from the Cassini mission shows that this assumption is quite realistic for an integration time of 60 s \citep{Asmar2005}. A 0.003 mm/s at 1000 s integration time could be assumed for the two-way range rate measurement accuracy. For white noise, this scales to 0.012 mm/s at 60 s, and 0.03 mm/s at 10 s. These expectations for Juno's Ka band Doppler system are supported by the analysis of Cassini cruise data \citep{Mariotti2013}. The simulated value of 0.020 mm/s at 60 s is larger than the expected performance of a Ka-Ka two-way coherent radio link in order to account for some variability with tracking condition (at low Sun-Earth-Probe angles the plasma noise dominates and the stability of the link is reduced, see \citealp{Iess2014}).

The residuals in case \textit{A} do not show obvious signatures of Jupiter's oscillations, although a closer inspection of orbit \#10 residuals indicate some effect near perijove. Indeed, in some passages, depending on the phase of the modes, the oscillations combine to produce perturbations too weak to be observed in Juno Ka-band Doppler data. On the opposite, in some passages the perturbation is large enough to produce a signature that cannot be absorbed by other static coefficients.

Even if the residual Doppler signatures are small in case \textit{A}, the acoustic modes still produce a small bias in the estimate of the zonal coefficients. $J_2$ and $J_3$ appear to be the most affected, with estimation error (truth minus estimate) within 5-sigma of the formal uncertainty (simulations without the acoustic modes show that the error on the harmonics is nominally within 3-sigma).

The situation changes dramatically if the energy of the modes is large (case \textit{B}). Residuals cannot be fitted to the noise level and a clear, unabsorbed, signal remains after the fit. This implies that Jupiter's acoustic perturbations on the gravity field are large enough to produce a detectable signal in Juno Doppler data and a large bias on the harmonic coefficients. In addition, as the estimated harmonic coefficients do not correlate with Jupiter's modes (essentially because they produce accelerations at different time scales), increasing the degree and order of the estimated gravity field cannot absorb acoustic mode signatures in Doppler data. Indeed, the signatures are still large even when a full 20x20 tesseral field plus $J_{21}-J_{30}$ zonal field is estimated.

The Love numbers $k_2$ and $k_3$ are other important quantities whose estimate can be adversely affected by acoustic oscillations. $k_2$ and $k_3$ control Jupiter's response to the tides raised by the Galilean satellites Io, Europa and Ganymede (Callisto's tide is too small to be revealed). As the three tides have very different periods, Juno can in principle measure the tidal response at different frequencies. This measurement can be spoiled by acoustic modes.

In fact, the estimation of Love numbers $k_2$ and $k_3$ appears critical. When no oscillations are included and in the case of shallow winds, the Love numbers can be retrieved with a formal accuracy of: $\sigma_{k_2}=1.5 \times 10^{-3}$, $\sigma_{k_3}=3.7 \times 10^{-3}$ (assuming $k_2=0.5$, $k_3=0.2$, the relative uncertainty is respectively 0.3\% and 1.8\%) and the estimation error is within 2-sigma, thus enabling a good retrieval of Jupiter’s tidal response. (The accuracies are comparable for the deep winds case.) When acoustic oscillations are included, the estimation error on the Love numbers is several times larger than the corresponding uncertainty. The results are summarized in table \ref{tab:2}. In case \textit{A}, the estimation of the degree 3 Love number is much degraded, with an estimation error of about 8 times its formal uncertainty. The estimate of $k_2$ is within 4-sigma from truth. These values will obviously change for different realizations of the modes. Although a full Monte Carlo simulation would be more appropriate to determine the statistics of the estimation errors when systematic effects are present (which is the case), in this initial work we aim only to assess if the presence of the oscillations is appreciable. Over the next two years Juno and new ground instruments \citep{Schmider2013} will indicate if acoustic modes need to be accounted for in the dynamical model and the data analysis. 

In case \textit{B} the effect of the acoustic modes on the retrieval of Jupiter's Love numbers is much larger. The estimation error on the $k_2$ and $k_3$ is about 300 times the associated formal uncertainty, invalidating the measurement of Jupiter's response to tides. This behavior is not in itself surprising because the perturbations generated by Jupiter's internal acoustic modes mimic the effect of the tides during Juno passages at perijove. Acoustic oscillations are seen as a dynamical noise that cannot be effectively separated from the accelerations due to the tidal bulges. Indeed, although the time scales of the two effects are certainly different (Io's orbital period is about 1.769 days; degree 2 fundamental mode oscillate every 140 minutes), Juno's highest sensitivity period to Jupiter gravity field is limited in time to a few hours near perijove, thus limiting our ability to distinguish the two effects.

\begin{table}
\centering
\begin{tabular}{lcccc}
\hline
  & Reference &     Formal    & Estimation error & Estimation error \\
  &    value      & uncertainty & in case \textit{A} & in case \textit{B} \\
\hline
$k_2$ & 0.5 & $1.5 \times 10^{-3}$ & $6.3 \times 10^{-3}$ $(\cong 4 \sigma)$ & $5.2 \times 10^{-1}$ $(> 300 \sigma)$ \\
$k_3$ & 0.2 & $3.7 \times 10^{-3}$ & $2.7 \times 10^{-2}$ $(\cong 8 \sigma)$ & $1.1 \times 10^{-0}$ $(> 300 \sigma)$ \\
\hline
\end{tabular}
\caption{Love numbers estimation error to be compared with its formal uncertainty, in the two cases.}
\label{tab:2}
\end{table}

The second scenario (strong modes) requires a new approach to the data analysis. In order to avoid biasing the harmonic coefficients and fit the data close to the noise level, the acoustic mode amplitudes, frequencies and phases need to be estimated. Of course this would be required only for those modes with a larger impact on the data. Otherwise, not only Juno Doppler data could not be fitted, but the solution for the static gravity field would also be untruthful. In fact, the static gravity field solution obtained in this scenario shows an estimation error for each degree that is one to two orders of magnitude larger than the corresponding expected formal uncertainty.


\section{Conclusions}
\label{sec:5}

In this paper we evaluated the impact of Jupiter's acoustic modes to the Juno gravity experiment. The effect appears as acoustic modes propagating through the planet perturb its density profile, thus changing the external gravity field (whose determination is the main goal of the Juno radio science experiment).

\cite{Gaulme2011} reported the detection of global oscillations of Jupiter from ground-based observations with a maximum radial velocity of 49 cm/s @ 1213 $\mu$Hz for Jupiter's upper troposphere, providing a first, though non-exhaustive, indication of the mode amplitudes. However, the signal detected from ground provides information on modes with radial order $n \sim 4$ to $11$ rather than on the lower radial order modes (in particular the $n=0$ ones) which are supposed to be responsible for the largest gravity perturbations.

We thus chose to focus on two different scenarios for the mode amplitudes. Case \textit{A} is based on a velocity-to-frequency profile derived from the observed values from \cite{Gaulme2011}. For the unobserved f-modes ($n=0$) we have assumed an amplitude of one hundred times smaller than that of observed modes, a ratio similar to what has been measured for the Sun. Case \textit{B} represents a more energetic scenario, with f-modes having a similar amplitude as the one of the detected p-modes. 

For both cases, we performed simulations with the orbit determination code to be used for the analysis of Juno's radio science data, including also the effect of acoustic modes. The results of the simulations indicate that one cannot rule out a significant effect of Jupiter's modes in Juno Doppler data. In particular, Juno observables appear to be more sensitive to low degree, low radial order modes. In the two cases analyzed, Juno Doppler data shows signature coming from Jupiter's oscillations (see fig. \ref{fig:6}), but the main actors are different in the two cases. In the strong mode scenario \textit{B} the signatures are mainly produced by the fundamental modes, which induce the largest perturbation (see fig. \ref{fig:4}). When the mode amplitude is large, a static field expansion is inadequate to absorb the effect of Jupiter's oscillation on Doppler data, so that large signatures appear in the post-fit residuals. These signatures would be an indication of an inadequate dynamical model, and of the unreliability of the gravity solution. The estimation of Jupiter Love numbers $k_2$ and $k_3$ would be impossible. In this case a new data analysis approach would be required, unescapably entailing the estimation of the mode amplitude, frequency and phase. 

In case \textit{A} the post-fit residuals do not show a clear signature of acoustic modes, although indiscernible evidence is present in some perijove passages, depending on whether the modes interact constructively. In that case, the low $n$ modes have much lower amplitudes (hundred times less than peak amplitude modes), producing only small perturbations on Jupiter gravitational field. For the low f-mode amplitudes assumed within case \textit{A}, it would be difficult to extract information on the modes from Juno Doppler data. On the other hand, the effect on the estimation of Jupiter's static field and Love numbers will be minimal (although biases at the level of a few sigma would still be possible, especially for the Love numbers).

To conclude, Juno can potentially provide useful information on low degree, low radial order acoustic oscillations modes, if their amplitudes are large enough (above $\sim$ 1 cm/s for $n=0$ modes). If the modes are weak (at the level assumed in case \textit{A}), Juno data will not show any significant dynamical perturbation associated with Jupiter global modes. In this case, it will be possible to derive an upper limit on Jupiter acoustic mode amplitudes. If the modes are strong (as assumed in case \textit{B} of our analysis), the dynamical model used to fit the Doppler data must include the effect of the modes and the estimation of their amplitude, frequency and phase for all relevant degree and order. A comparison between the Juno's measurements and the new ground-based observations with the newly developed, highly sensitive, JIVE and JOVIAL instruments (respectively, Jovian Interiors Velocimetry Experiment, and Jovian Oscillations through radial Velocimetry ImAging observations at several Longitudes) would be very important \citep{Schmider2013}.


\section*{Acknowledgments}

The work of D.D and L.I. has been supported in part by the Italian Space Agency. T.G. acknowledges support from CNES. The authors would like to thank D. J. Stevenson for pointing out the relevance of normal modes in Juno's measurements and for stimulating discussions on this topic.



\bibliography{mybib}

\end{document}